# Size-Dependent Band-Tail Localization in Oxide Semiconductors Revealed by Direct Density-of-States Mapping

Chang Niu[†], Kisoo Nam[†], Aravindh Shankar, Jian-Yu Lin, Sanjeev Khare, Sumi Lee, Pramey Upadhyaya, and Peide D. Ye*

*Elmore Family School of Electrical and Computer Engineering and Birck Nanotechnology Center, Purdue University, West Lafayette, IN 47907, United States.*

†These authors contributed equally to this work: Chang Niu, Kisoo Nam

*Correspondence and requests for materials should be addressed to P. D. Y. (yep@purdue.edu)

**Disorder-induced localization is expected to become increasingly important as amorphous oxide semiconductor transistors are scaled toward low-dimensional channels, yet the electronic states responsible for this transport regime remain difficult to resolve experimentally. Here, we use a lock-in-based electric-field penetration technique to directly map the effective density of states (DOS) in In-based oxide semiconductor thin-film transistors (TFTs). The extracted quantum capacitance, carrier density, and chemical potential reveal a disorder-dominated transport regime in which band-tail states are not merely passive traps, but become screening-active and partially transport-active. Geometry-dependent DOS mapping shows an exponential suppression of the effective DOS with channel length, demonstrating size-dependent band-tail localization and providing a microscopic origin for a distinct localization-induced threshold-voltage roll-off mechanism. Temperature-dependent measurements show that the disorder-dominated DOS is strongly suppressed at low temperatures, while extended diffusive states remain nearly unchanged, confirming the localization origin. By tuning film thickness, $O_2$ annealing, and In/Ga/Zn composition, we further demonstrate systematic suppression of disorder and effective-DOS localization. This work establishes direct DOS mapping as a device-level probe of localization physics and provides a pathway for engineering disorder in low-dimensional oxide semiconductor electronics.**

Oxide semiconductors (OSs) have emerged as important channel materials for next-generation thin-film electronics because they combine wide bandgaps,[1] low off-state leakage,[2] high electron mobility,[3,4] and compatibility with back-end-of-line (BEOL) processing[5–7]. These properties make OS thin-film transistors (TFTs) attractive for displays, monolithic three-dimensional (3D) integration, and heterogeneous electronic systems. However, in contrast to crystalline 3D bulk semiconductors such as Si, Ge, and III–V compounds, low-dimensional semiconducting channels are strongly affected by structural disorder, interface roughness, defects, and potential fluctuations.[8] As OS devices are scaled toward ultrathin and ultrashort channels, these disorder effects become increasingly important,[9] not only as sources of trap states and reliability degradation, but also as fundamental factors governing carrier transport through disorder-induced localization.[8]

A long-standing challenge in semiconductor transistors is understanding the physical role of band-tail states,[10–14] which have been recognized as key electronic states governing subthreshold swing ($SS$),[15] threshold voltage ($V_{th}$), mobility, and bias-stress stability. These states are conventionally regarded as static trap-like distributions that affect transistor characteristics without directly participating in transport. Such a picture, however, becomes incomplete in scaled, low-dimensional,[8,16] or cryogenic transistors,[17,18] where disorder plays an amplified role in carrier motion. In this regime, a substantial fraction of band-tail states can become screening-active and transport-active. When the device length approaches or exceeds the localization length ($\xi$), electronic states near the band edge become spatially localized and deviate from conventional diffusive transport. The resulting effective density of states (DOS) is therefore no longer a purely material-intrinsic quantity but can depend on device size or measured temperature through localization. This size-dependent population of transport-active band-tail states gives rise to scaling behavior that cannot be captured by contact resistance, electrostatics, or conventional trap models alone.

In this work, we emphasize the importance of this third transport regime, beyond well-studied diffusive and ballistic transport regimes, the disorder-dominated regime in which disorder effects are no longer negligible but actively govern carrier screening and transport. Directly identifying the band-tail states requires an experimental probe beyond standard current-voltage analysis. Using a lock-in-based electric-field penetration technique (eFPT), we directly map the DOS in In-based oxide semiconductor TFTs and reveal that localization-induced band-tail states dominate carrier transport, leading to the breakdown of conventional scaling and a new and previously unrecognized $V_{th}$ roll-off mechanism. By systematically analyzing size-, temperature-, film-thickness-, annealing-, and In-concentration-dependent transfer characteristics and DOS maps, we establish a comprehensive framework for describing and experimentally quantifying disorder-induced electron localization. Our work provides a pathway for probing and engineering disorders as well as localized electronic states in low-dimensional transistors.

**Disorder-Dominated Transport Regime**

Scaling is the central principle that has driven the continuous performance improvement of semiconductor technologies. In conventional transistors, this scaling relies on Ohm's law, where channel resistance increases linearly with channel length. This linear resistance scaling, however, is valid only in the diffusive transport regime. More generally, carrier transport in a transistor channel is governed by the relationship among three characteristic lengths: the channel physical length $L$, the mean free path $l$, and the localization length $\xi$, which defines the spatial extent of electronic wave functions. As shown in **Figure 1a**, when $L < l$, the device enters the ballistic regime, where carriers traverse the channel with negligible scattering and the two-terminal resistance becomes nearly independent of channel length, approaching the Landauer quantum limit $R = h/(Me^2)$, where $M$ is the number of conducting modes including spin and valley degeneracies. When $l < L < \xi$, transport is diffusive and follows Ohmic scaling, $R \propto L$.

However, when $L > \xi$, carrier localization dominates transport. In this disorder-dominated regime, relevant for OS and other low-dimensional field-effect transistors (FETs), the resistance increases exponentially with channel length, leading to a fundamental breakdown of conventional scaling.[8] **Figure 1b** illustrates the scaling relationship between the total device resistance $R$ and channel length $L$. Ohmic scaling breaks down in both the ballistic and disorder-dominated regimes, but in opposite ways: ballistic transport produces weak length dependence, whereas disorder-dominated transport leads to an exponential increase in resistance with $L$. Despite its importance for scaled low-dimensional transistors, this third transport regime has received limited attention.

In an ideal crystalline semiconducting material in the two-dimensional (2D) limit, the DOS follows a step-function profile, with $D_{2D} = \frac{g_s g_v m^*}{2\pi\hbar^2}$ (1), where $g_s$, $g_v$, and $m^*$ are the spin degeneracy, valley degeneracy, and effective mass, respectively. In the presence of disorder, potential fluctuations spatially localize a fraction of electronic states near the band edge, preventing them from participating in transport as extended diffusive states. This localization leads to the formation of band-tail states, as illustrated in **Figure 1c**. According to their energy and transport activity, these band-tail states can be divided into three regimes: deep trap states located inside the bandgap, shallow localized tail states, and disorder-dominated tail states that are sensitive to device size and measured temperature. At higher energies, extended diffusive states are recovered and contribute to conventional transport.

## Direct DOS Mapping by Electric-Field Penetration Technique

Although standard transfer characteristics, capacitance-voltage measurements,[16,19] and charge-pumping[20] techniques can reveal threshold-voltage shifts, SS degradation, non-Ohmic scaling, or trap distributions, they do not directly resolve the electronic DOS as a function of energy in an operating transistor. Spectroscopic techniques such as scanning tunneling spectroscopy[21,22] and ultraviolet photoelectron spectroscopy[1] provide material-level DOS information, but are generally not directly connected to the electrostatic screening and transport response of fabricated devices. Direct DOS mapping in transistor geometries is therefore required to quantitatively identify band-tail states and correlate them with disorder-induced localization. Here, we develop a lock-in-based electric-field penetration technique to directly map the DOS in thin-body semiconductor transistors by probing the screening response of electronic states.[23–25] In the dual-gate geometry shown in **Figure 2a**, an AC excitation ($V_{AC}$) applied to the back gate generates a penetration field through the semiconductor channel, which is governed by the channel quantum capacitance $C_q$ and detected as a top-gate AC current $I_{TG}$. By sweeping the DC back-gate bias $V_{DC}$, the carrier screening response is monitored from depletion to accumulation. As described by the equivalent circuit in **Figure 2b**, normalizing $I_{TG}$ to the depletion-limit current $I_{TG}^0$, where channel screening is negligible, enables extraction of carrier density $n_s$, chemical potential $\mu$, $C_q$, and the DOS. This method directly probes the electronic states responsible for gate-field screening and carrier transport. **Figure 2c** shows the oxide semiconductor TFT structure used for DOS mapping, with the fabrication process detailed in **Figure S1**.

A standard eFPT measurement of a 2.8-nm-thick $InO_x$ TFT with $W_{ch} = L_{ch} = 20\ \mu m$ at room temperature is shown in **Figure 2d**. The $I_{TG}$ reaches a plateau in the full depletion regime, where the channel provides negligible screening, and decreases in the accumulation regime due to enhanced carrier screening. The corresponding $C_q$ and DOS as a function of $V_{BG}$ are extracted using:

$$\frac{I_{TG}}{I_{TG}^0} = \frac{C_{TG} + C_{BG}}{C_{TG} + C_{BG} + C_q};\ C_q = e^2 \frac{dn_s}{d\mu} = e^2 D(E) \qquad (2)$$

Here, $C_{TG}$ and $C_{BG}$ are the top- and back-gate geometric capacitances. The carrier density induced by the back gate is obtained by taking the full depletion regime as the zero-density reference and integrating the quantum-capacitance-limited gate response:

$$n_s(V_{BG}) = \frac{1}{e}\int_{V_{dep}}^{V_{BG}} \frac{C_q(V_{BG})C_{BG}}{C_q(V_{BG}) + C_{TG}+C_{BG}} dV \qquad (3)$$

The chemical potential is then obtained from the quantum-capacitance relation as:

$$d\mu(V_{BG}) = \frac{e}{C_q} dn_s \qquad (4)$$

when $\mu$ is expressed in eV. The slow increase in conductivity during transistor turn-on is frequently observed in OS TFTs and 2D-material FETs, yet it is often described phenomenologically through mobility degradation, trap filling, Schottky barrier effect or subthreshold behavior. By comparing the transfer curve (**Figure 2e**), we show that this gradual turn-on originates from a disorder-dominated regime, where band-tail states progressively become transport-active. Only after the DOS reaches a nearly constant value does the device transition into the extended diffusive regime. This result provides a microscopic explanation for the broadened turn-on characteristics and indicates that disorder-induced band-tail localization can additionally increase the required operating voltage[26,27] and limit switching steepness. Finally, by plotting the DOS on a logarithmic scale as a function of chemical potential (**Figure 2f** and **Figure S2**), four distinct electronic regimes are resolved: deep trap states with negligible current conduction, shallow band-tail states associated with the subthreshold regime, disorder-dominated band-tail states corresponding to the slow increase in conductivity, and extended diffusive states where conventional transport is recovered. Among these regimes, the disorder-dominated band-tail regime has received limited attention in previous studies, despite its important role in carrier transport and device scaling with atomically thin 2D transistors particularly.[4,8]

## Geometric- and Temperature- Dependent Effective DOS

Besides the broadened turn-on characteristics, another distinct feature of the disorder-dominated transport regime is the nonlinear geometric-dependent conductance scaling arising from strong carrier localization. **Figure 3a** shows the transfer characteristics of $InO_x$ TFTs with channel dimensions ranging from 10 to 50 μm and a fixed aspect ratio of $W_{ch}: L_{ch} = 1:1$. The low contact resistance of $InO_x$ and the relatively long channel lengths strongly suppress contact-related contributions.[28] In addition, negligible hysteresis indicates stable transistor operation and minimizes the influence of trap-related effects. A pronounced $V_{th}$ shift is observed, with saturation emerging at channel lengths of around 40 μm. This behavior is in sharp contrast to conventional Si transistors, for which nearly identical transfer characteristics would be expected over this length scale. We identify it as a distinct $V_{th}$ roll-off mechanism arising from disorder-induced localization, as supported by the evidence below.

The corresponding eFPT penetration-current response and effective DOS extracted from devices of different geometries on the same $InO_x$ film are shown in **Figures 3b** and **3c**. The chemical potential for different channel sizes is referenced to zero gate bias, assuming a common intrinsic chemical-potential reference for devices fabricated from the same film. The extracted effective DOS reflects the electronic states that actively participate in screening and transport. Transport in OS TFTs can be understood as the coexistence of localization-dominated and extended diffusive contributions, with the localization component determined by the competition between the energy-dependent localization length $\xi(E)$ and the physical channel length $L$. At higher energies, closer to the conduction band, $\xi(E)$ increases; once $\xi(E) > L$, these states become fully transport-active. In the extended-diffusive-state-dominated limit, the channel-length dependence vanishes, and the system recovers conventional diffusive transport with linear scaling. The slightly sharper DOS increase observed in the 50 μm device indicates that only a small fraction of localized states persists in this regime.

To quantify the geometry dependence, the effective DOS at fixed energy levels is plotted against channel length in **Figure 3d**. The effective DOS decreases exponentially with increasing channel length, a hallmark of strong localization, and follows:

$$D_{eff}(E,L) = D_0(E)\exp\left[-\frac{L}{\xi(E)}\right] \quad (5)$$

Here, $D_0(E)$ is the DOS at short-channel limit. The slopes from the linear fits in **Figure 3d**, together with the extracted localization lengths, are summarized in **Figure S3**. Consistently, an independent localization length extracted from conductance scaling, shown in **Figure S4**, falls within the same order of magnitude. This agreement provides direct evidence that the geometry-dependent effective DOS originates from carrier localization.

The DOS is commonly treated as an intrinsic, temperature-independent quantity, with temperature affecting transport mainly through Fermi–Dirac occupation. In disorder-dominated transistors, however, this picture breaks down because localization changes the population of electronic states that can screen the gate field and participate in transport. As temperature decreases, band-tail states become increasingly localized, leading to a reduction in the effective, transport-active DOS. This behavior is evident in the temperature-dependent resistance of a 2.8-nm-thick $InO_x$ TFT shown in **Figure 4a**, where strong carrier freeze-out appears at low temperatures. The corresponding effective DOS maps extracted from the same device are shown in **Figure 4b**. Four regimes are again resolved, consistent with **Figure 2f**. Among them, only the disorder-dominated regime exhibits a pronounced temperature dependence, whereas the extended diffusive states remain nearly unchanged. Similar effective DOS localization is observed in a 10×10 $\mu m^2$ device shown in **Figure S5**. The shallow band-tail region becomes steeper upon cooling, reflecting reduced thermal broadening. The temperature dependence is quantified in **Figure 4c**, where the effective DOS is plotted at selected chemical potentials. States closer to the band edge show much stronger temperature dependence, confirming their localization-dominated character. Geometry-dependent effective DOS maps measured at 35 K are shown in **Figure 4d**, demonstrating that size-dependent localization remains robust at cryogenic temperatures.

Geometry- and temperature-dependent characterizations reveal a distinct localization-induced $V_{th}$ roll-off mechanism. The saturation behavior observed at long channel lengths indicates the coexistence of extended diffusive transport and localization-dominated transport. Unlike the diffusive component, the localization-dominated contribution is highly sensitive to channel geometry and temperature, as electron localization strongly limits the population of transport-active states.

**Suppressing Disorder in Oxide Semiconductors**

With the identification of the disorder-dominated transport regime, engineering disorder in transistors becomes increasingly important, particularly in amorphous oxide semiconductors and low-dimensional transistors such as 2D-material or nanosheet devices with strong thickness confinement, where band-tail transport is no longer negligible. Disorder can originate from multiple sources, including structural disorder, interface quality, defect density, and thickness roughness. Here, we focus on disorder associated with the semiconductor channel itself by evaluating the effects of film thickness, In concentration, and $O_2$ annealing.

We first performed geometry-dependent DOS mapping on 2.0-nm-thick $InO_x$ films (**Figure 5a**) and 3.8-nm-thick $InO_x$ films (**Figure S6**). As the film thickness increases and the device gradually evolves from a strong two-dimensional limit toward a more three-dimensional transport geometry, the degree of the impact from the disorder is reduced.[8] $O_2$ annealing provides another effective approach for suppressing disorders. High-temperature annealing can partially crystallize the film and reduce structural disorder, while the $O_2$ ambient passivates oxygen-vacancy-related defects. As shown in **Figure 5b**, $O_2$ annealing induces a positive $V_{th}$ shift and significantly suppresses the channel-length-dependent $V_{th}$ shift. Consistently, the channel-length dependence of the effective DOS is strongly reduced after annealing. The disorder-dominated regime highlighted in **Figure 5b** shows only a weak channel-length dependence, providing clear evidence that disorder-induced localization has been suppressed. Below this regime, the device operates in the subthreshold region, whereas above it, extended diffusive states dominate and the device recovers the conventional linear scaling. In addition to thickness engineering and $O_2$ annealing, doping $InO_x$ with Ga and Zn also reduces the degree of disorder by stabilizing the oxide semiconductor network. **Figure 5c** and **Figure S7** shows geometry-dependent characterization of an IGZO film with a comparable thickness and an In:Ga:Zn elemental ratio of 5:1:1 and 7:1:1. Compared with pure $InO_x$, the IGZO device exhibits a reduced geometry-dependent $V_{th}$ shift and weaker effective-DOS localization. The geometric scaling parameters extracted from process-, composition-, and thickness-dependent DOS measurements are summarized in **Figure 5d**. These parameters quantify the disorder-induced localization of the effective DOS and reveal a clear trend: disorders in In-based oxide semiconductors can be reduced by increasing film thickness, lowering In concentration through Ga/Zn doping, and applying $O_2$ annealing. These results provide a systematic strategy for engineering disorder and localized band-tail states in future low-dimensional semiconductor technologies.

## Conclusion

In summary, we directly map the effective DOS in In-based oxide semiconductor TFTs using a lock-in-based eFPT and reveal a disorder-dominated transport regime governed by band-tail localization. The extracted effective DOS exhibits a strong geometry dependence and follows an exponential decay with channel length, demonstrating that transport-active band-tail states are limited by a finite device-size comparable localization length. This localization-induced DOS suppression provides a microscopic origin for a distinct $V_{th}$ roll-off mechanism beyond conventional short-channel effects, contact resistance, and static trap models. Temperature-dependent measurements further confirm the localization origin. By tuning film thickness, $O_2$ annealing, and In/Ga/Zn composition, we further demonstrate the systematic suppression of the impact of disorders and the reduction of effective-DOS localization. These findings establish eFPT as a powerful approach for resolving transport-relevant electronic states and provide design principles for engineering of disorders in low-dimensional oxide semiconductor transistors.

## Methods

### Device Fabrication

High-resistivity Si substrates (>10,000 Ω·cm) with 300 nm $SiO_2$ were used to minimize parasitic capacitance between the substrate, metal pads, and transistor channel. A 10 nm $Al_2O_3$ adhesion layer was deposited by thermal atomic layer deposition (ALD) using trimethylaluminum (TMA) and oxygen plasma. A 35 nm Ni bottom gate was then deposited by electron-beam evaporation. $HfO_2$ bottom-gate dielectrics with different thicknesses were deposited at 200 °C by plasma-enhanced ALD (PEALD) using tetrakis(dimethylamido)hafnium (TDMAHf). $InO_x$ and IGZO channel layers were deposited at 225 °C by thermal ALD. Trimethylindium (TMIn), diethylzinc (DEZ), $Ga_2(NMe_2)_6$, and $H_2O$ were used as In, Zn, Ga, and O precursors, respectively. Devices with different channel dimensions were patterned using AZ1518 photoresist and a Heidelberg MLA150 maskless aligner. Source/drain contacts were formed by electron-beam evaporation of 35 nm Ni, followed by channel isolation using a Plasma-Therm Apex SLR ICP-RIE system. A 12 nm $HfO_2$ top-gate dielectric was deposited at 150 °C by PEALD, and a 50 nm Ni top gate was subsequently formed by electron-beam evaporation. Selected devices were annealed in $O_2$ at 250 or 300 °C for 1 min using a Jipelec rapid thermal annealing system.

**Device characterization**

Transfer characteristics were measured using a Keysight B1500 semiconductor parameter analyzer. Gate-dielectric capacitances were measured using an Agilent E4980A LCR meter. Electric-field penetration measurements were performed using an SR830 lock-in amplifier together with a Keithley 2450 source meter. Temperature-dependent measurements from 7 to 295 K were conducted in a Lake Shore CRX-VF cryogenic probe station.

**Acknowledgements**

P.D.Y. was supported in part by the U.S. National Science Foundation under Award No. 2425498 with industry partners as specified in the Future of Semiconductors (FuSe2) program. P.D.Y. was also supported by Semiconductor Research Corporation Global Research Program and Samsung Electronics, Inc. (Grant Nos. IO250514-12841-01).

**Author Contributions**

C.N. and K.N. contributed equally to this work. P.D.Y. supervised the project. C.N. and K.N. conceived and designed the experiments, performed the electrical measurements, and analyzed the experimental data. C.N., K.N., J.-Y.L., and S.L. fabricated the devices. A.S. and S.K. performed the theoretical calculations and analyses under the supervision of P.U. P.D.Y., C.N., and K.N. wrote the manuscript with contributions from all authors.

**Competing financial interests**

The authors declare no competing financial interests.

**Supporting Information**

Additional details for fabrication flow, localization length extraction, and DOS mapping of different films are in the supplementary information.

**Corresponding Author**

* Peide D. Ye (E-mail: yep@purdue.edu)

# Figures

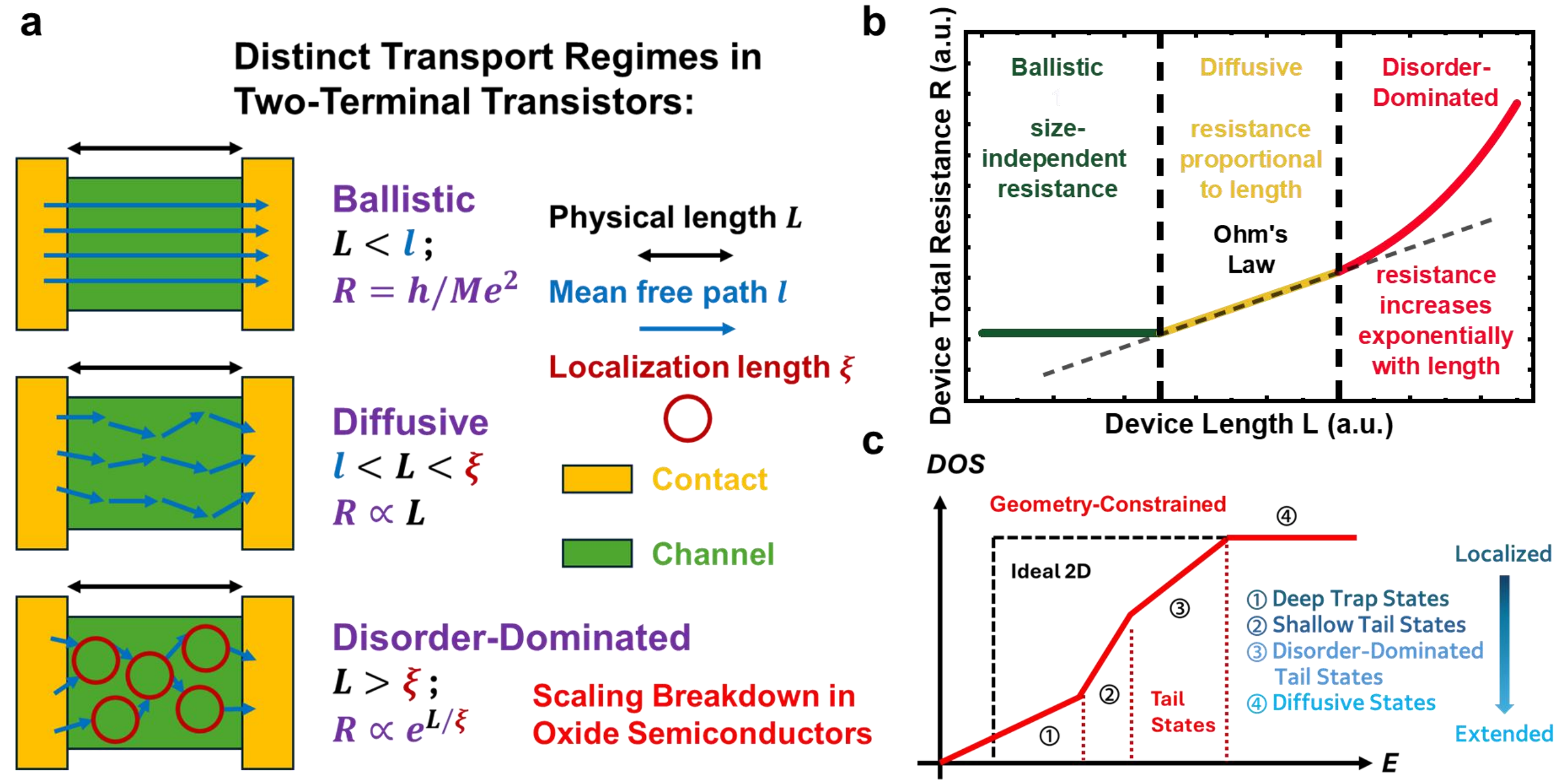

**Figure 1. Disorder-Dominated Transport Regime. a**, Schematic of ballistic, diffusive, and disorder-dominated transport regimes in transistor channels, defined by the relationship among the channel length $L$, mean free path $l$, and localization length $\xi$. The resistance is length-independent in the ballistic regime ($L < l$), scales linearly with $L$ in the diffusive regime ($l < L < \xi$), and increases exponentially with $L$ in the disorder-dominated regime ($L > \xi$). **b**, Scaling of the total device resistance $R$ with device geometry, illustrating the breakdown of Ohmic scaling in both the ballistic and disorder-dominated regimes. **c**, Schematic illustration of the effective DOS involved in transport in disordered channel materials. In the ideal 2D limit, the DOS follows a step-function profile corresponding to diffusive states. In the presence of disorder, band-tail states develop, including deep trap states, shallow tail states, and disorder-dominated tail states.

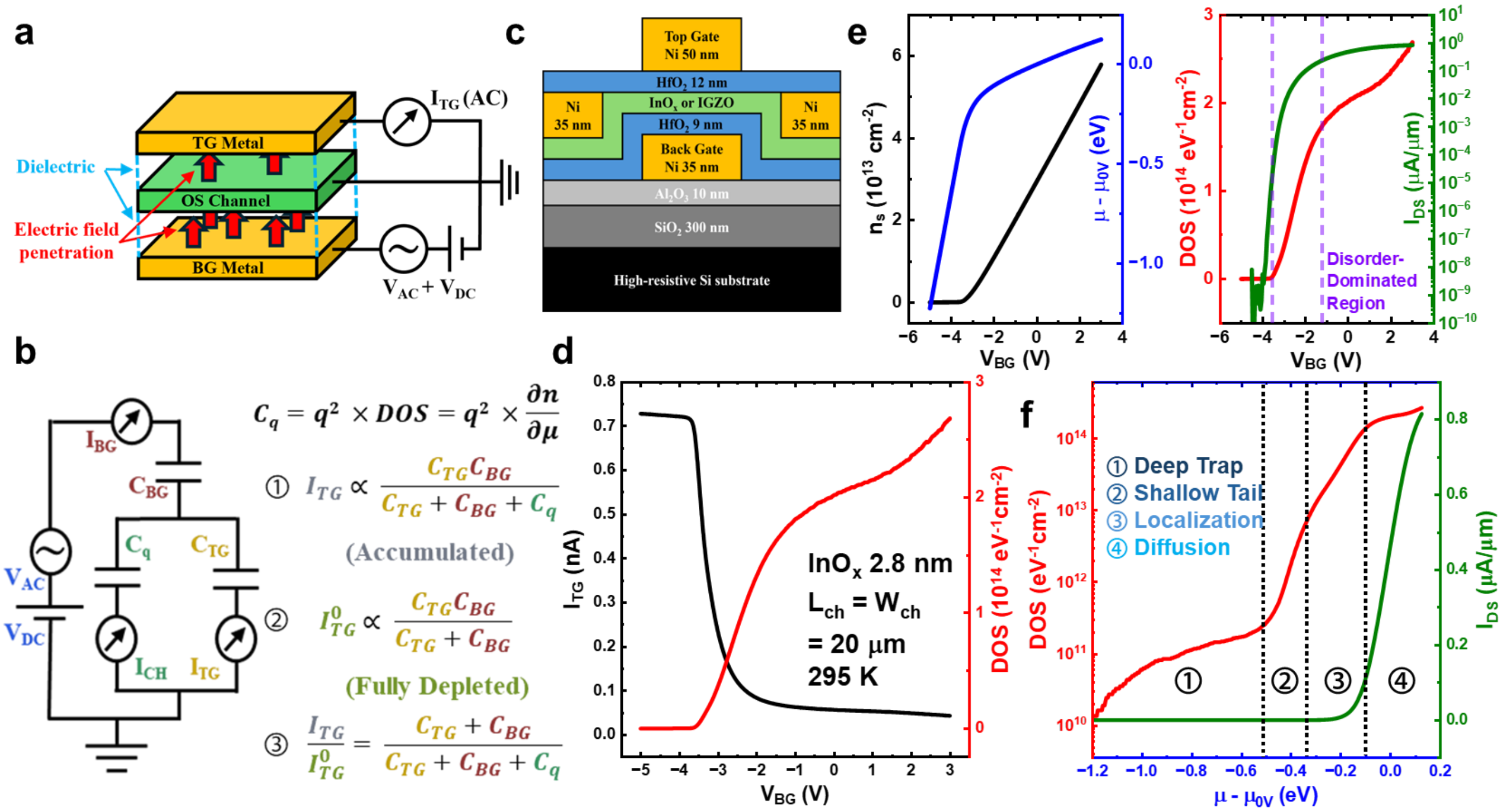

**Figure 2. Direct DOS Mapping by the Electric-Field Penetration Technique (eFPT).** **a**, Schematic of the eFPT in a dual-gate $InO_x$ device. A small AC excitation superimposed on the back-gate voltage $V_{BG}$ generates a penetration field that is detected at the top gate as $I_{TG}$. **b**, Equivalent circuit of the eFPT measurement. The top-gate current depends on the $C_q$, enabling extraction of the DOS from the depletion-to-accumulation response. **c**, Schematic of the dual-gate $InO_x$ TFT used for DOS mapping. **d**, Measured $I_{TG}$ and extracted DOS as a function of $V_{BG}$ for a 2.8-nm-thick $InO_x$ TFT with $W_{ch} = L_{ch} = 20\ \mu m$. **e**, Carrier density, chemical potential, DOS, and source-drain current as a function of $V_{BG}$, resolving the depletion, disorder-dominated and accumulation regimes. **f**, DOS and source–drain current plotted against chemical potential, revealing four electronic regimes from deep trap states to extended diffusive states.

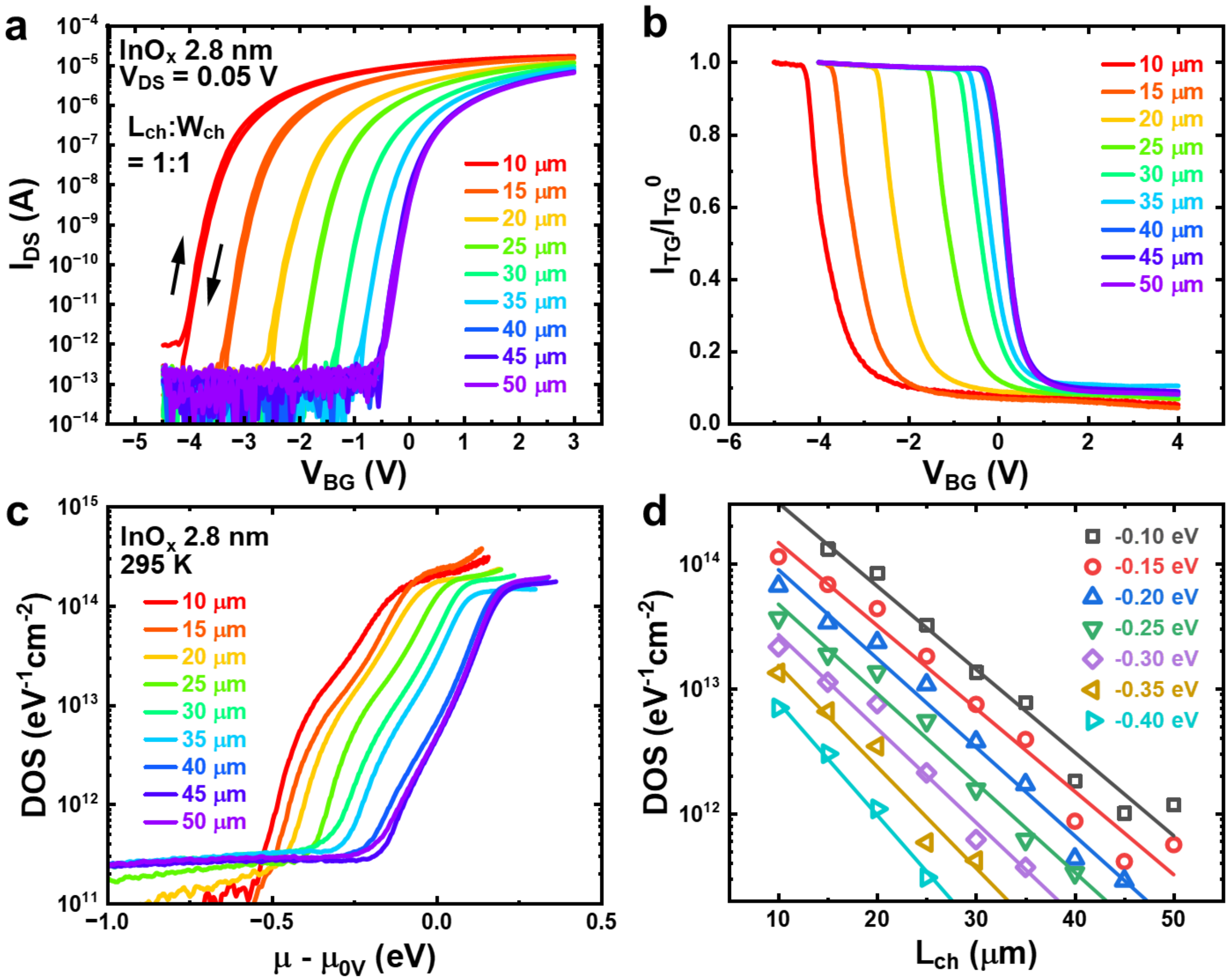

**Figure 3. Size-Dependent Electrical Transport and Effective DOS. a**, Transfer characteristics of 2.8-nm-thick $InO_x$ TFTs with $W_{ch} = L_{ch}$, showing negligible hysteresis and a pronounced size-dependent $V_{th}$ shift. **b**, Normalized top-gate current $I_{TG}/I_{TG}^0$ measured from same devices by eFPT. **c**, Effective DOS extracted from devices with different channel sizes on the same film, revealing size-dependent transport-active band-tail states. **d**, Effective DOS plotted as a function of channel length at fixed chemical potential, showing exponential suppression with increasing $L_{ch}$, a signature of localization-limited transport.

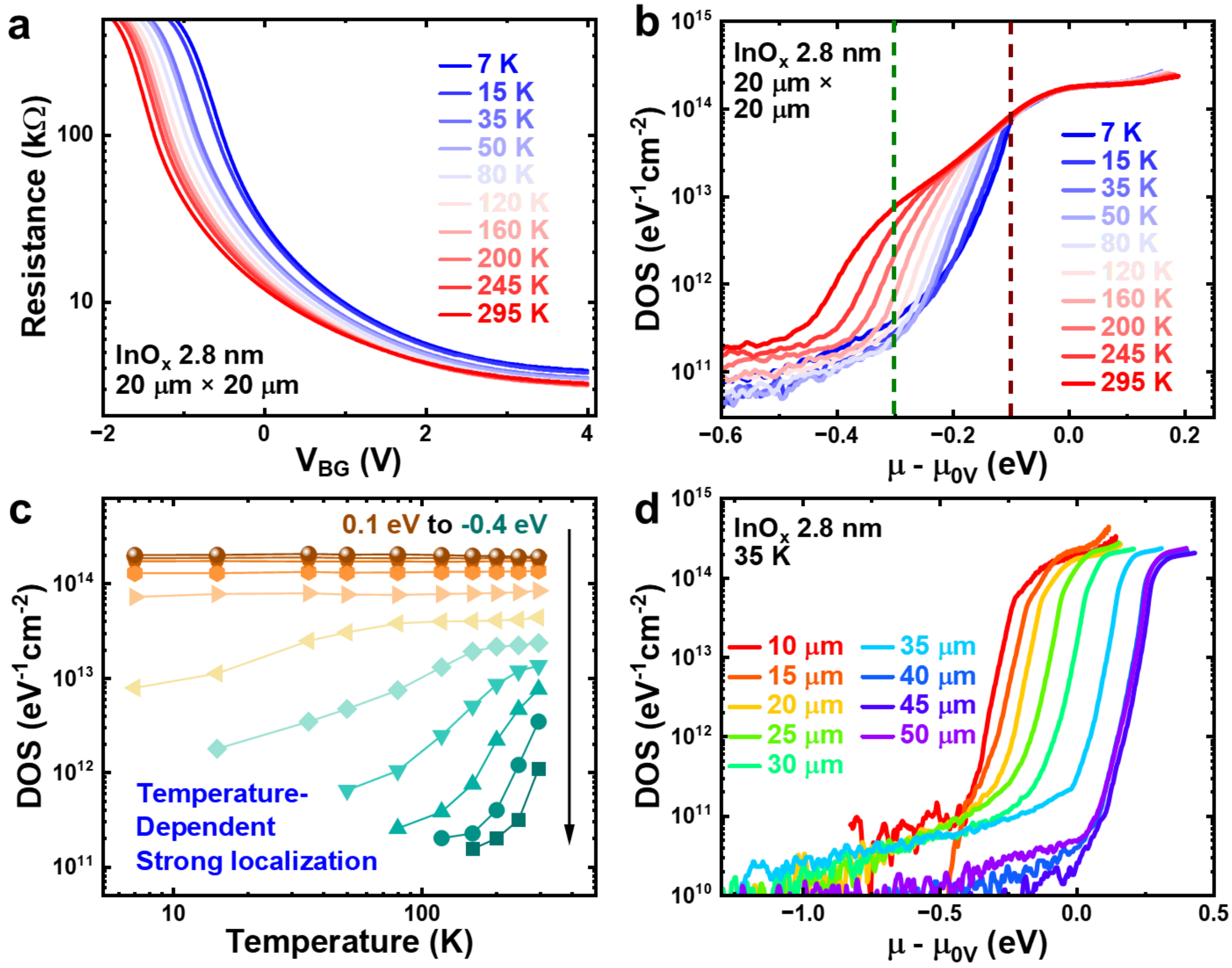

**Figure 4. Temperature-Dependent Electrical Transport and Effective DOS.** **a**, Resistance as a function of $V_{BG}$ measured at different temperatures. **b**, Temperature-dependent DOS maps extracted from the same devices. The disorder-dominated regime exhibits strong temperature-enhanced localization, whereas the diffusive states remain largely unchanged. **c**, DOS as a function of temperature at different chemical potentials, showing the evolution from diffusive transport to strongly localized states. **d**, Size-dependent DOS maps measured at 35 K, showing that both channel geometry and temperature are key factors governing localization-dominated transport.

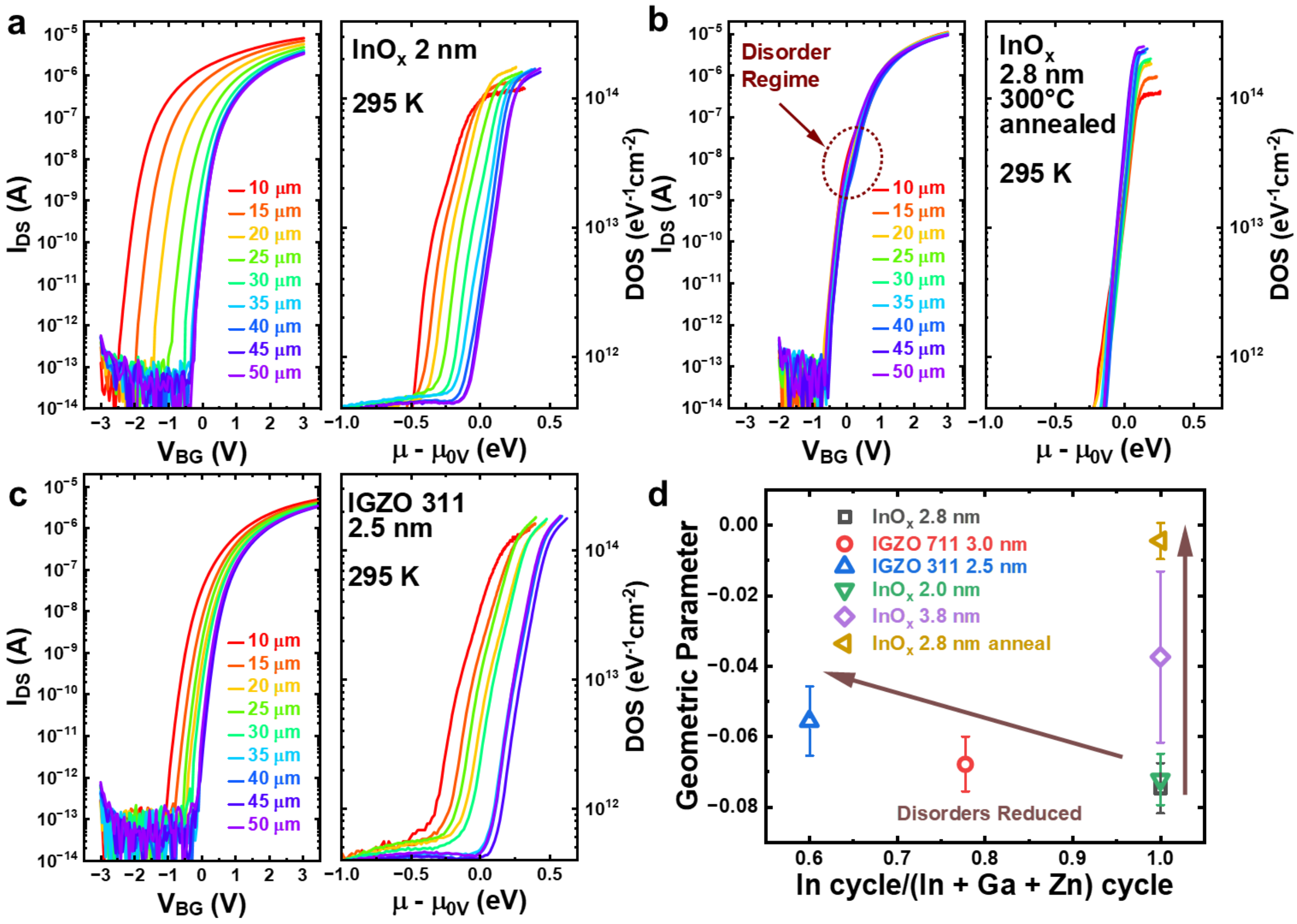

**Figure 5. Suppressed Disorder in Oxide Semiconductors. a**, Transfer characteristics and effective DOS maps of size-dependent 2.0-nm-thick $InO_x$ TFTs. **b**, Transfer characteristics and effective DOS maps of size-dependent 2.8-nm-thick $InO_x$ TFTs after 300 °C $O_2$ annealing, showing minimal channel-length dependence. **c**, Transfer characteristics and effective DOS maps of size-dependent 2.5-nm-thick IGZO 311 TFTs, revealing weaker disorder effects. **d**, Geometric parameters extracted from the slope of the DOS decay as a function of channel length under different film thicknesses, compositions and process conditions, showing that the degree of disorder decreases with lower In concentration, increased thickness and $O_2$ annealing.

# Supplementary Information for:

# Size-Dependent Band-Tail Localization in Oxide Semiconductors Revealed by Direct Density-of-States Mapping

Chang Niu[†], Kisoo Nam[†], Aravindh Shankar, Jian-Yu Lin, Sanjeev Khare, Sumi Lee, Pramey Upadhyaya, and Peide D. Ye*

*Elmore Family School of Electrical and Computer Engineering and Birck Nanotechnology Center, Purdue University, West Lafayette, IN 47907, United States.*

†These authors contributed equally to this work: Chang Niu, Kisoo Nam

*Correspondence and requests for materials should be addressed to P. D. Y. (yep@purdue.edu)

**List of contents:**

**Supplementary figures:**

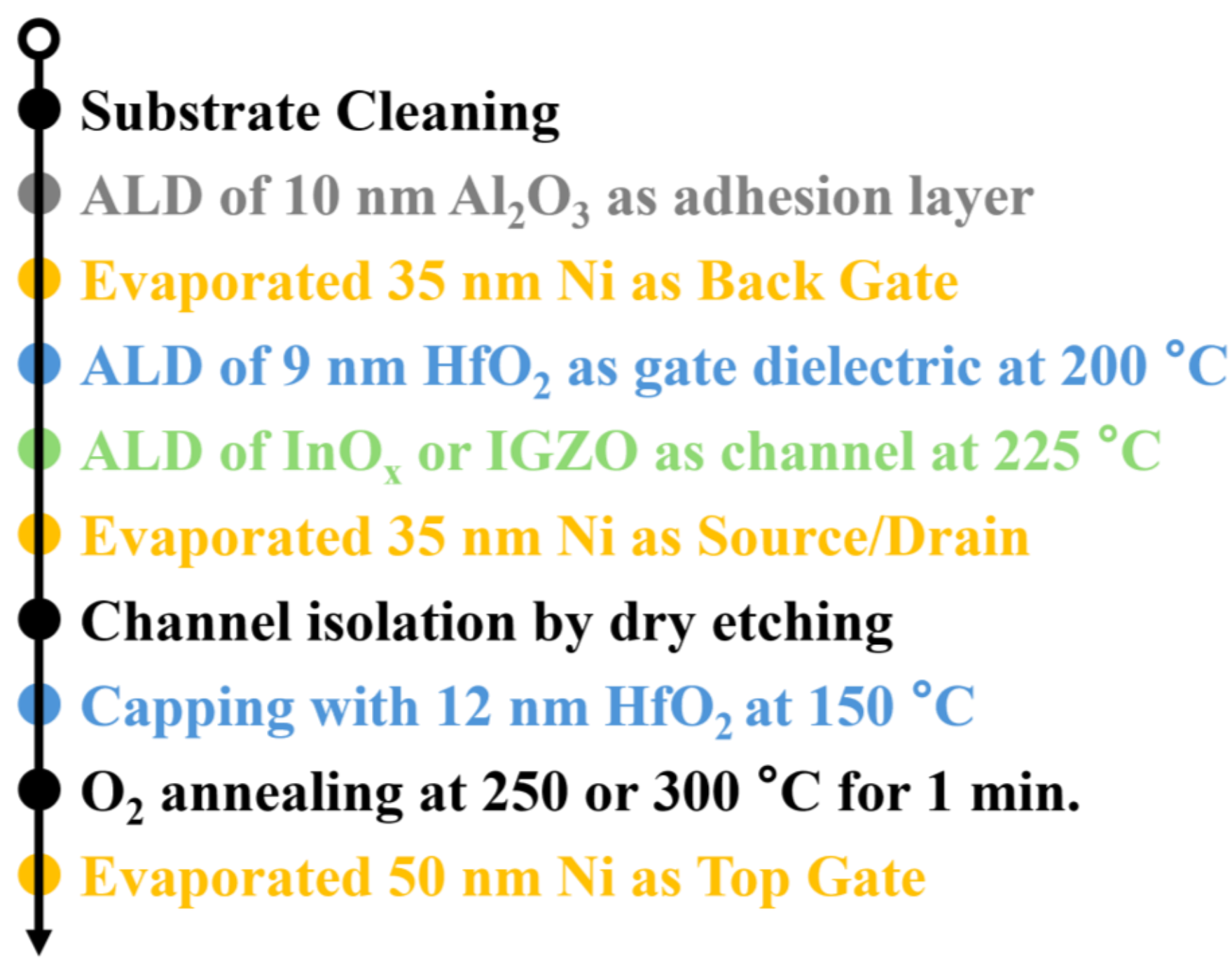

**Figure S1. Fabrication flow of dual-gate OS TFTs.** Schematic fabrication process of the dual-gate oxide semiconductor thin-film transistors used for electric-field penetration measurements. High-resistivity Si substrates were used to minimize parasitic capacitance from the substrate. The dual-gate structure enables detection of the penetration electric field across the semiconductor channel.

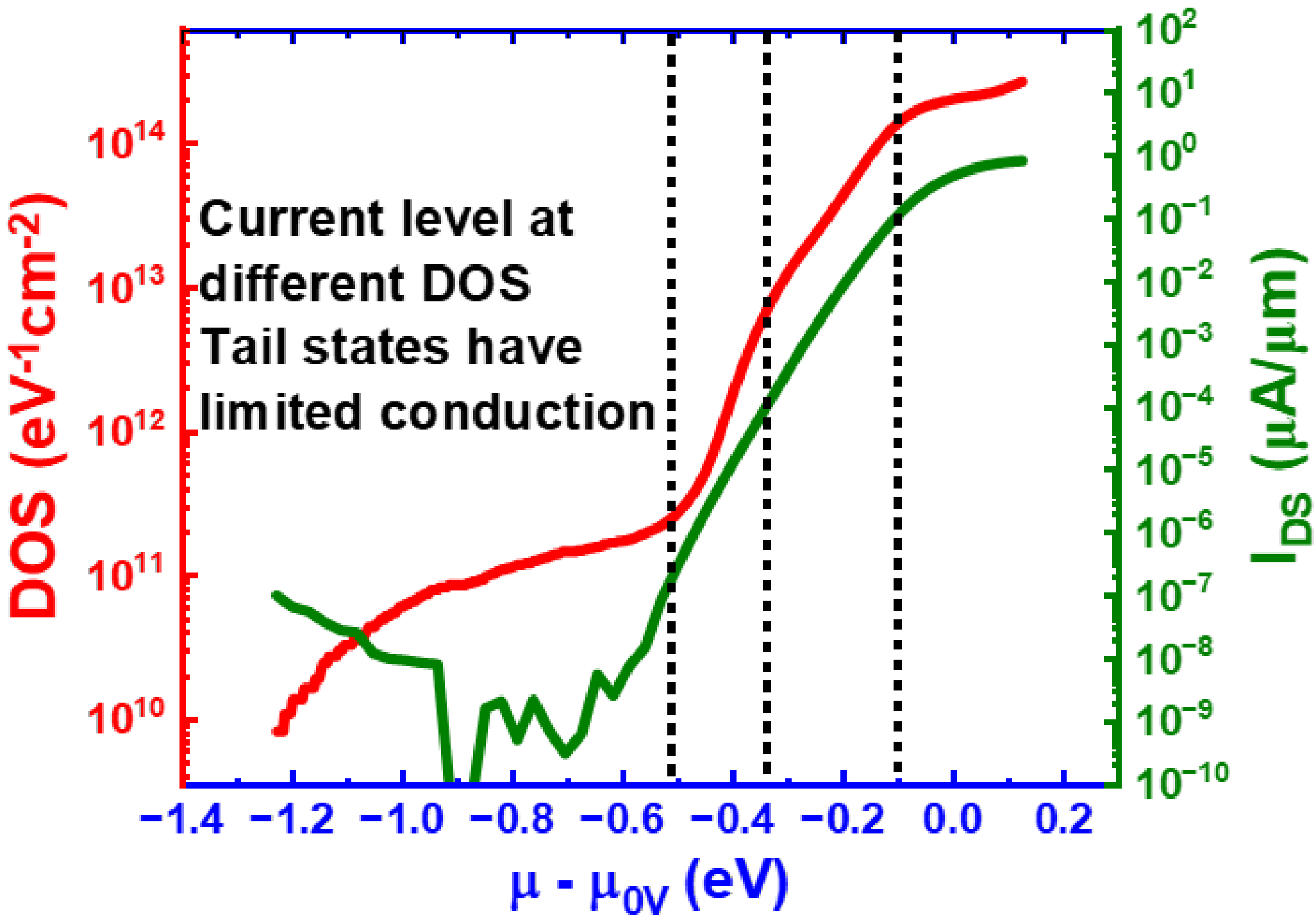

**Figure S2. Comparison between DOS mapping and source–drain current.** Effective DOS and source–drain current plotted as a function of chemical potential. By comparing the DOS map with the transfer characteristics, different transport regimes can be identified from both the electronic-state and conductivity perspectives.

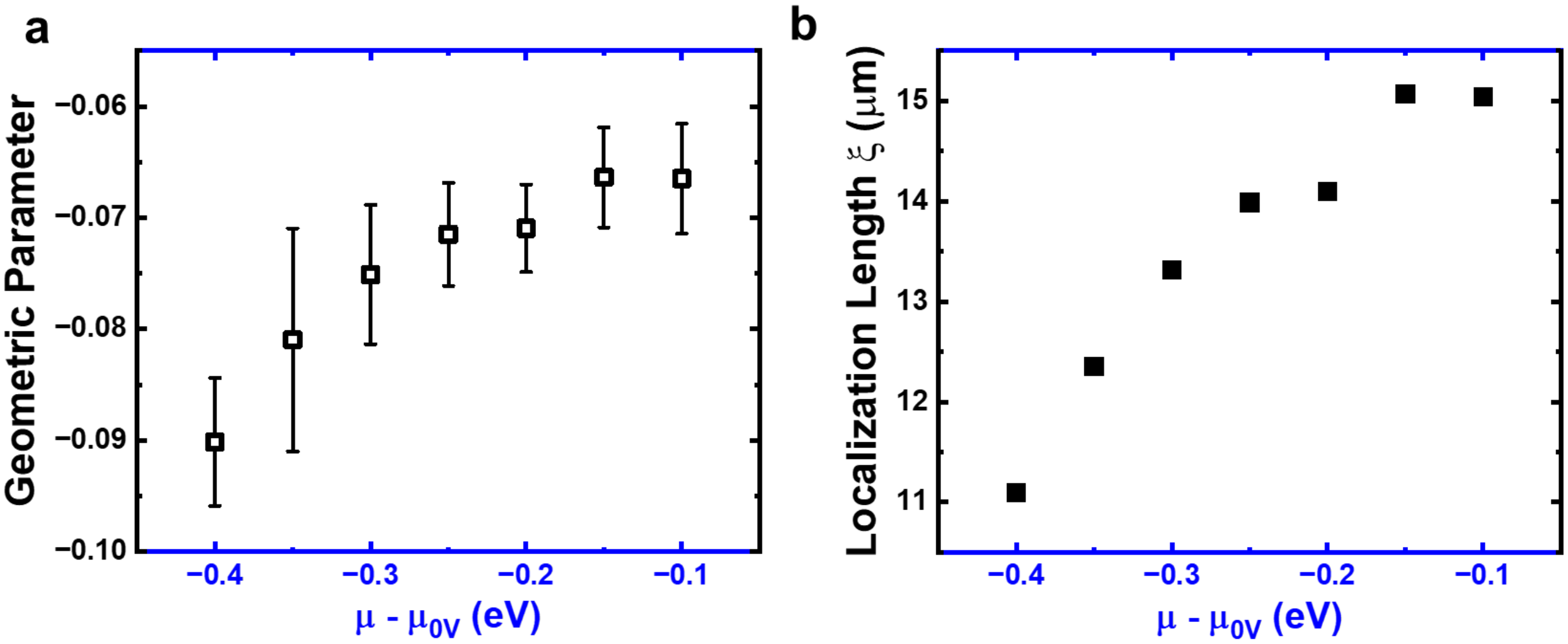

**Figure S3. Localization length extracted from effective DOS mapping.** Geometric scaling parameter characterizing the localization of the effective DOS and the corresponding localization length plotted as a function of chemical potential. The extracted localization length increases as the chemical potential approaches the extended-state regime, indicating weakened localization at higher energies.

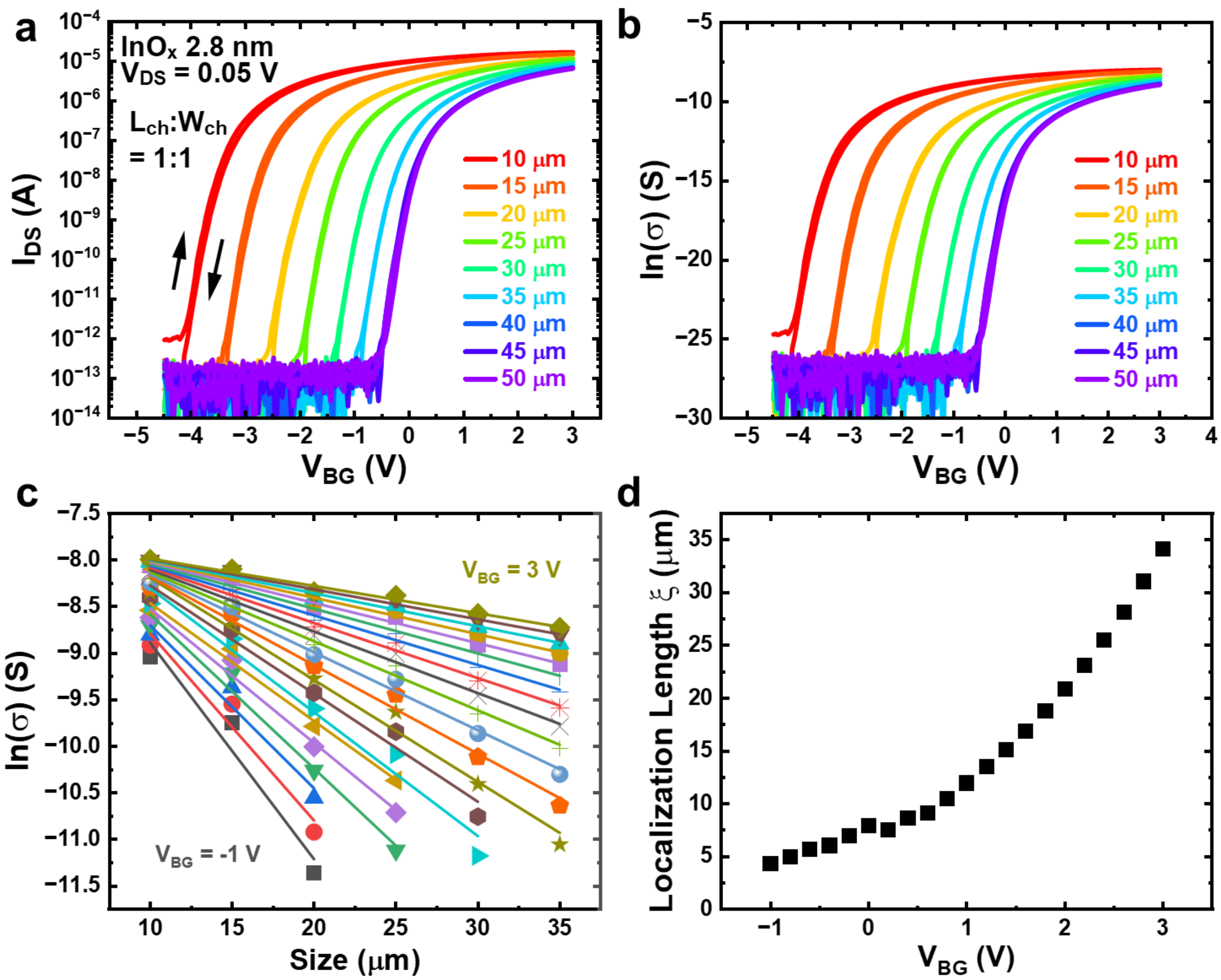

**Figure S4. Extraction of localization length from conductivity scaling. a**, Geometry-dependent transfer characteristics of devices fabricated from the same $InO_x$ film. **b**, Calculated conductivity as a function of back-gate voltage for devices with different geometries. **c**, Conductivity plotted as a function of channel size, showing exponential suppression with increasing device length. **d**, Localization length as a function of back-gate voltage extracted from the slopes in **c**.

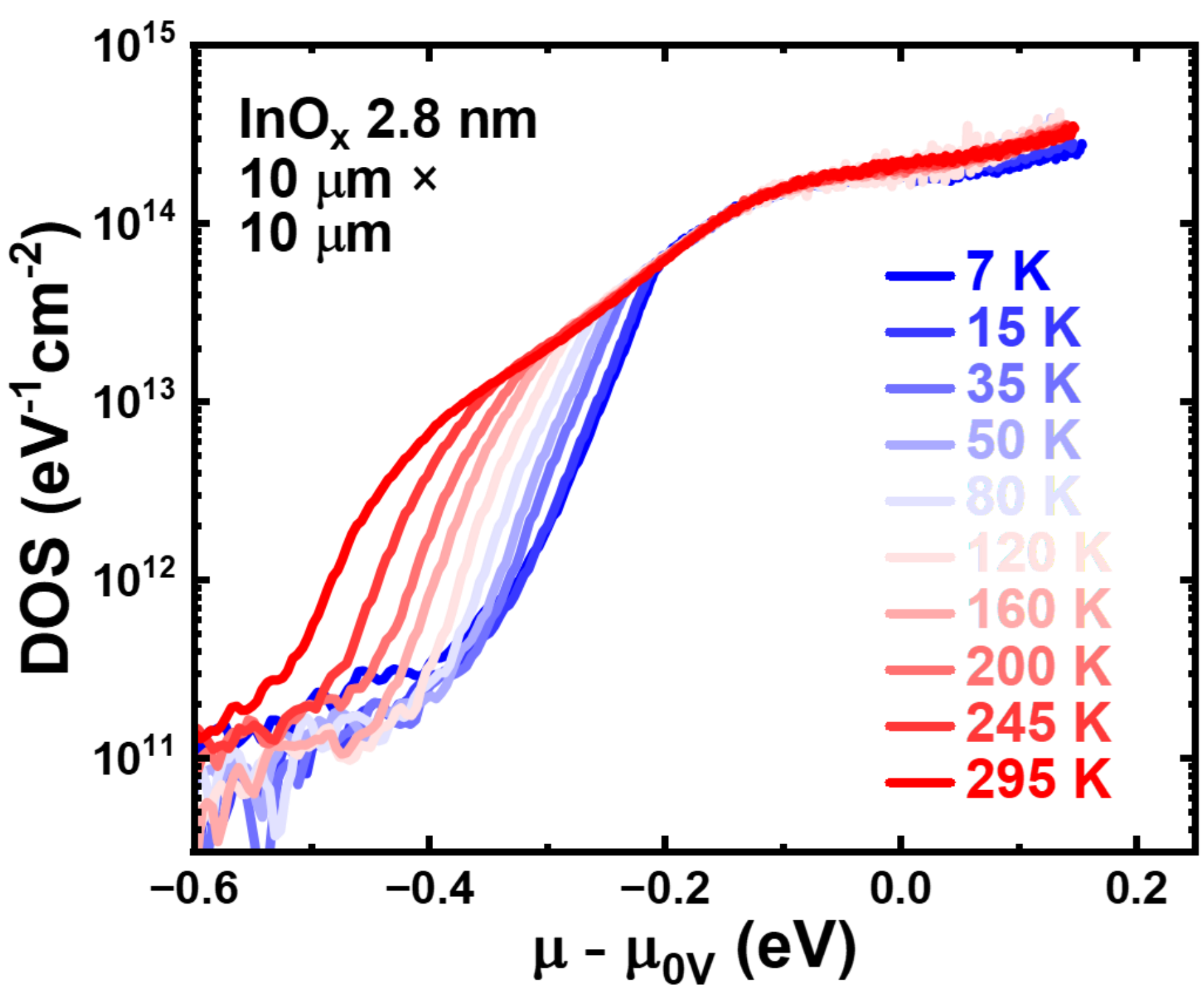

**Figure S5. Temperature-dependent evolution of band-tail states in another device.** Temperature-dependent effective DOS maps measured from another $InO_x$ TFT with smaller dimensions, showing the suppression of disorder-dominated band-tail states upon cooling and confirming the reproducibility of temperature-dependent localization behavior.

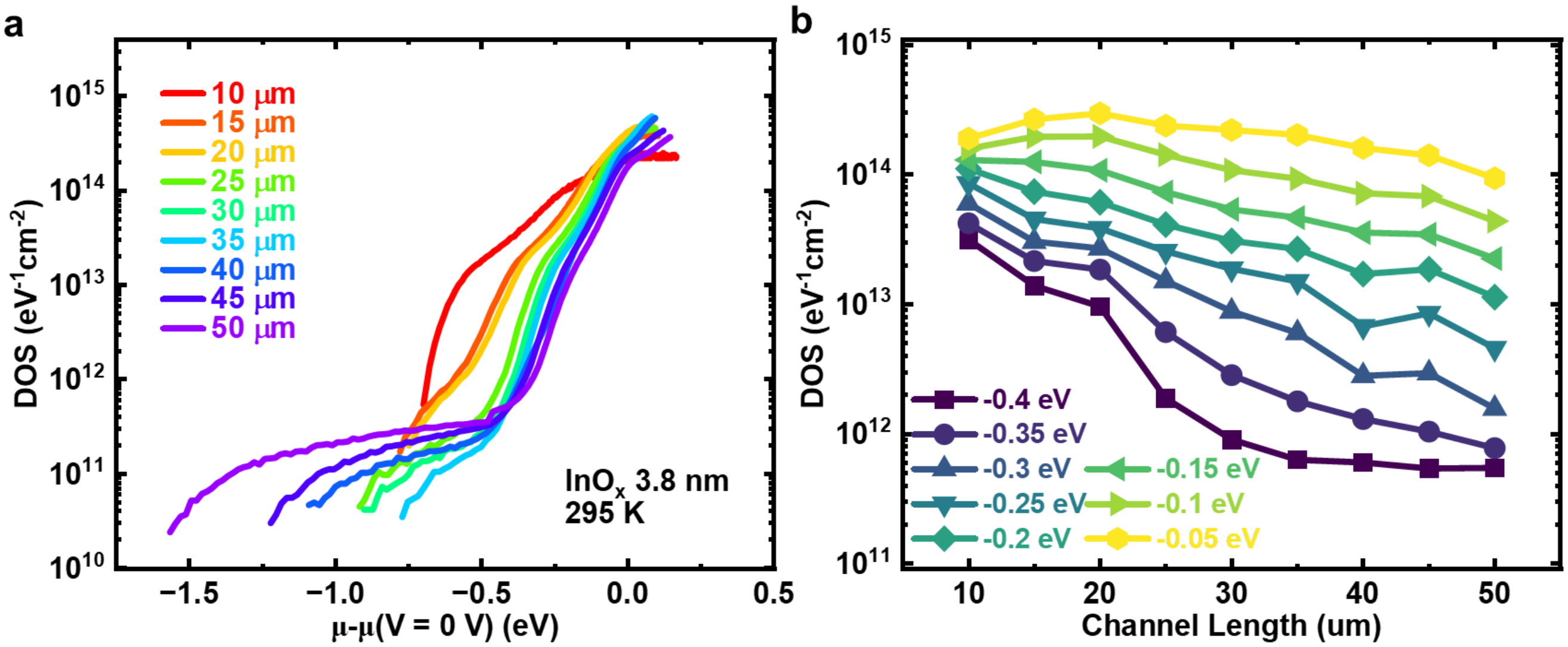

**Figure S6. Effective DOS maps of size-dependent 3.8-nm-thick $InO_x$ TFTs.** Effective DOS maps (**a**) and geometric-dependent localization (**b**).

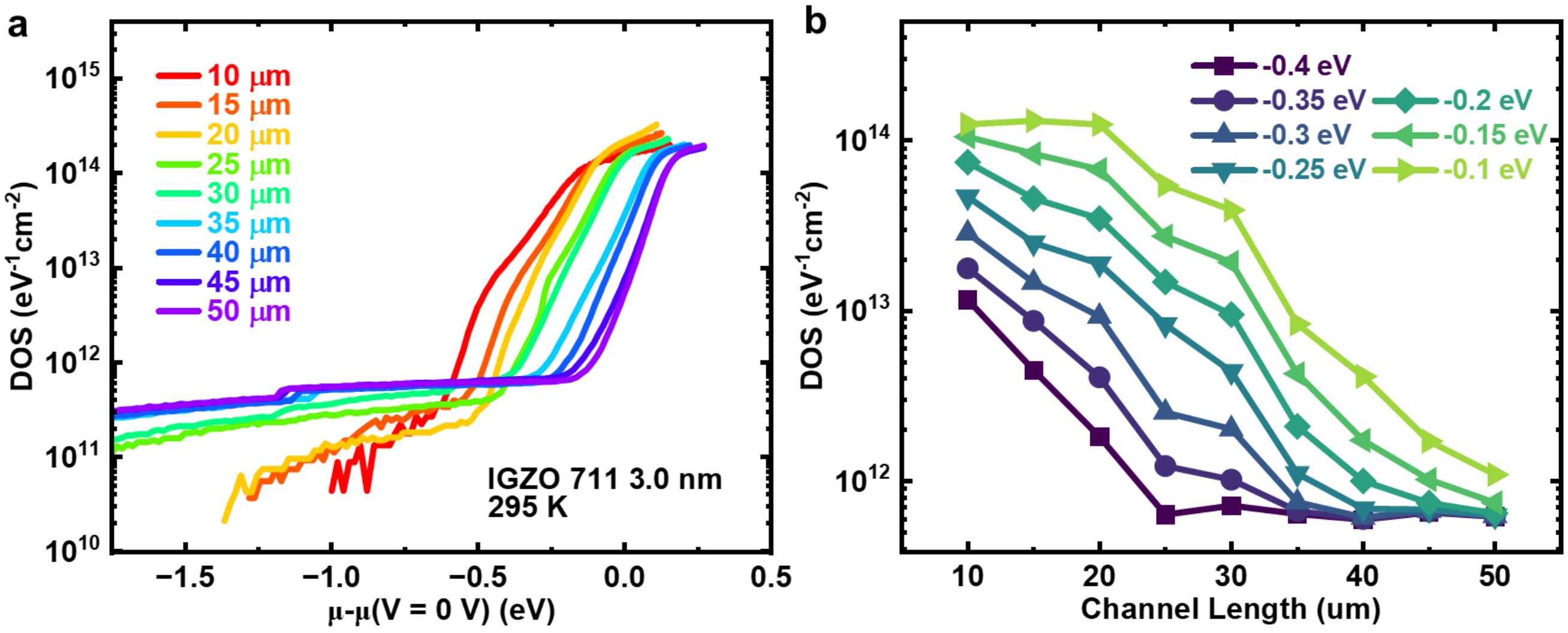

**Figure S7. Effective DOS maps of size-dependent 3.0-nm-thick 711 IGZO TFTs.** Effective DOS maps (**a**) and geometric-dependent localization (**b**).